\documentstyle[aps,prl,multicol,epsf]{revtex}

\begin{document}
\draft
\preprint{}

\title{Magnetoresistance of Granular Ferromagnets - Observation of a
Magnetic Proximity Effect?}

\author{A. Frydman and R.C. Dynes} 

\address{Department of Physics, University of California, San Diego,
La Jolla, CA 92093}

\maketitle

\begin{abstract}

We have observed a superparamagnetic to ferromagnetic transition in 
films of isolated Ni grains covered by non-magnetic 
overlayers. The magnetoresistance (MR) of the films was measured as a 
function of the overlayer thickness. 
Initially, the granular Ni films exhibited negative MR
curves peaked at H=0. As different materials were deposited onto the grains
hysteresis developed in the
MR. This behavior is ascribed to an increase of the typical domain size
due to magnetic coupling between grains. The strength of the inter-grain
coupling is found to correlate with the magnetic susceptibility of the overlayer material.
We discuss possible mechanisms for this coupling and
suggest that the data may reflect the existence of a magnetic
proximity effect (analogous to the well-known effect in superconductivity)
in which a ferromagnetic moment is induced in the metallic
non-magnetic medium.  

\end{abstract}

\pacs{PACS numbers: 75.50.Tt, 75.70.Pa, 73.40.Rw, 74.50.+r}

\begin{multicols}{2}

The proximity effect is a well known phenomenon in superconductivity
and has drawn a lot of interest both from the fundamental and the practical
points of view. In a high transmission superconductor-normal metal 
contact the superconductive wave-function varies smoothly across 
the interface causing a suppression of the pair amplitude
in the superconductor  and an enhancement of superconductivity on the 
normal side. An analogue 
in magnetism, in which the magnetization varies smoothly across
a ferromagnet-non magnetic metal interface, has been considered theoretically
 \cite{zukerman,cox,tersoff,white}. Experimentally, the observation of a 
magnetic proximity effect is much more challenging. In the first place
the coherence length of a typical ferromagnet, such as Ni or Fe,
is of the order of a few atomic spacings (this should be compared with
 $\xi_{s}$ of thousands
of {\AA} in a conventional superconductor) and the means to measure
the proximity effect are not as straightforward as in the superconducting
case. Furthermore, one has to be able to distinguish 
between a ``real'' proximity effect and other magnetic phenomena
such as magnetostatic interactions. 
One experimental approach in the past has been to study the suppression of magnetization in 
thin ferromagnets deposited on a normal metal substrate\cite{libermann,bergmann,gyorgy,Moodera}.
The first few monolayers were found to be magnetic ``dead layers''
exhibiting no ferromagnetic signal. In a different type of experiment Moodera et-al \cite{modera}
used spin-polarized tunneling measurements to show 
that a finite spin polarization of electrons persists in a Au film coupled to 
a Fe layer up to a thickness of a few tens of {\AA}. 
A question arises as to whether the spin 
polarization detection is indeed indicative of a ferromagnetic interaction inside the 
Au \cite{mersevey}.

In this paper we describe an experiment which is designed to probe a potential magnetic analogue to
the Josephson-like proximity coupling between superconductors across a SNS weak link. 
We measure the magnetoresistance (MR) of a granular magnetic film covered by various
non-magnetic overlayers and study the impact of these layers on the
magnetic coupling between the grains.

\begin{figure}[b]
\centerline{
\epsfxsize=90mm
\epsfbox{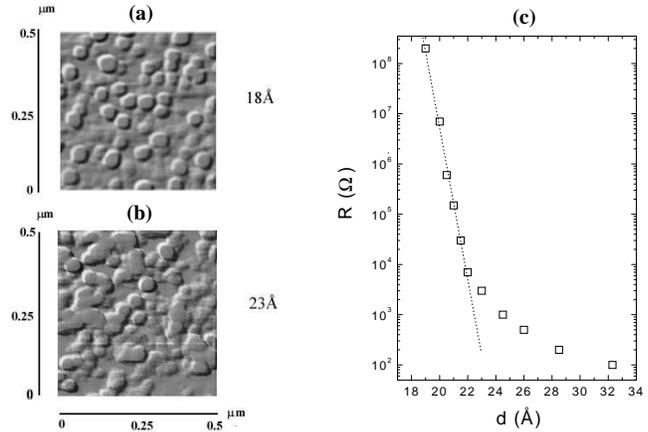}
}
\vspace{0mm}
\narrowtext
\caption{ Left: AFM micrographs of Ni films with nominal 
thickness of 18{\AA} and 23{\AA} evaporated on a room temperature
Si-SiO substrate. Line scans show that the average
hight of the grains is 35{\AA}. Right: Sheet resistance versus nominal
thickness of a quench-condensed Ni film.}
\end{figure}

The samples were prepared using the ``quench-condensation'' technique, 
i.e. evaporation
on a cryo-cooled substrate under UHV conditions within the measurement
apparatus. This method has a number of advantages for proximity effect 
experiments. First, the high vacuum
in the system gives rise to barrier-free interfaces between different evaporated 
materials. These are essential for proximity effect observation.
Furthermore, the low substrate temperature all but eliminates
inter-metallic diffusion, making the possibility of material alloying
highly unlikely. Another advantage of this fabrication method is that it
enables the evaporation of sequential layers of material   {\em in-situ}; 
thus one can study
the properties of a single sample as a function
of the amount of deposited material while keeping the sample at
low temperatures and in a UHV environment.

Our measurements were performed on thin layers of Ni quench-condensed onto
a Si-SiO substrate. The morphology of such samples is illustrated
in Figures 1a and 1b.  For thin enough films the structure contains isolated
grains with diameters of a few hundreds of {\AA} \cite{remark} and heights of 30-40 {\AA}. 
As more material is deposited, grains begin to coalesce 
with each other (Figure 1b). The average grain size thus increases and inter-grain spacing
decreases until, beyond a percolation threshold, the film becomes continuous. 
This behavior is also demonstrated by the dependence of resistance on nominal 
film thickness (Figure 1c). For small thicknesses the resistance drops exponentially
with thickness implying that the conduction mechanism is tunneling
or hopping between grains. For larger thicknesses the film becomes
continuous and the resistance crosses over to an ohmic 1/d dependence.

\begin{figure}[tb]
\centerline{
\epsfxsize=80mm
\epsfbox{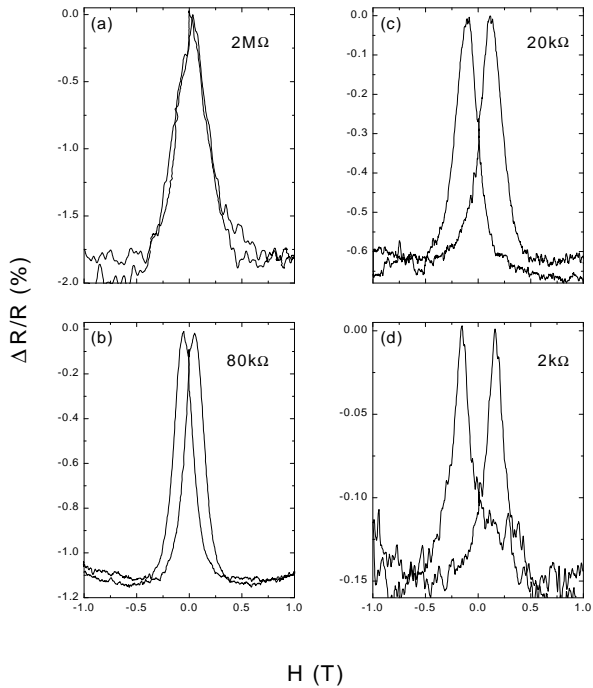}
}
\vspace{0mm}
\narrowtext
\caption{ Magnetoresistance curve at T=4.2K of a quench-condensed Ni
film for different steps of the evaporation. The field was swept 
from -1T to 1T and back. The nominal
film thicknesses were 20{\AA}(a), 21{\AA}(b), 21.8{\AA}(c) and 25{\AA}(d).}
\end{figure}

Figure 2 shows MR curves of a Ni film for different deposition steps.
The thinnest films exhibit curves which show 
a negative MR  centered at H=0 (Figure 2a). This behavior is similar
to that observed in other insulating granular ferromagnets prepared using
different fabrication techniques \cite{gittelman,gerber,yang,honda,sankar}
and is a result of spin dependent tunneling between grains which have
randomly oriented magnetic moments\cite{abeles}. Applying a field aligns these moments
causing a resistance decrease. These thin films consist of small grains which 
are isolated from
each other and are superparamagnetic at T=4K, thus,
when the field is removed, the thermal energy is large enough to randomize
the spin orientation and the MR curve does not exhibit hysteretic behavior.

As more material is deposited and the sheet resistance is reduced
hysteresis develops in the curve, resulting in two resistance peaks at finite
magnetic field. The position of the peaks
shifts towards larger fields as the film thickens (figures 2b, 2c and 2d).
Apparently, as the grains coalesce, the effective magnetic domain sizes become larger, 
the superparamagnetic blocking temperature rises and
the film exhibits ferromagnetic behavior at T=4K. Indeed, the temperature dependence
of these MR curves is consistent with well known relations for the superparamagnetic
transition\cite{future}. 
Adding material also causes the amplitude of the MR to decrease.
This is a result of the percolation
network for conductivity becoming denser, thus reducing the number of
tunneling events which occur between grains with different oriented
moments.
\begin{figure}[tb]
\centerline{
\epsfxsize=75mm
\epsfbox{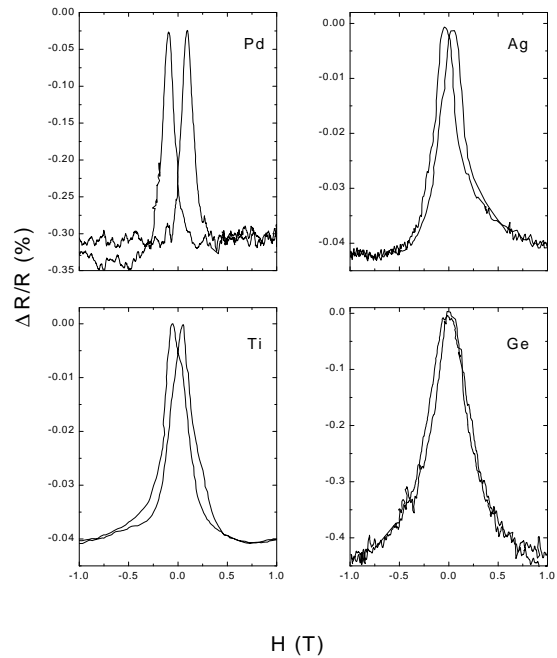}
}
\vspace{0mm}
\narrowtext
\caption{ MR curves at 4.2K of 20{\AA} thick films of granular Ni (R$\approx$4M$\Omega$) covered
by different overlayers. The initial MR curve for the bare Ni was similar
to that of figure 2a. For the Pd, Ti and Ag  the overlayer
thickness, d, is 6-7{\AA} and the sheet resistance is 2k$\Omega$.
For Ge d=65{\AA} and R=10K$\Omega$.}
\end{figure}

Having observed the above effect of adding magnetic material to a granular
magnetic film, we proceeded to study the effects of adding non-magnetic
materials to superparamagnetic grains. We prepared films of isolated Ni grains
(20{\AA} nominal thickness) which showed no hysteresis in the MR curve,
and we added overlayers of Pd, Ti, Ag, or Ge {\em in-situ}.
Figure 3
shows the MR curve of the samples in which an overlayer of 6{\AA} (or 65{\AA}
in the Ge case)  was added to the 20{\AA} Ni film.
Despite the fact that Pd, Ti and Ag overlayers are non ferromagnetic,
their presence gives rise to a hysteresis in the MR. Such a hysteresis is
indicative of coupling between magnetic grains which were originally 
isolated. The coupling strength is different for the different overlayer materials.

\begin{figure}[tb]
\centerline{
\epsfxsize=80mm
\epsfbox{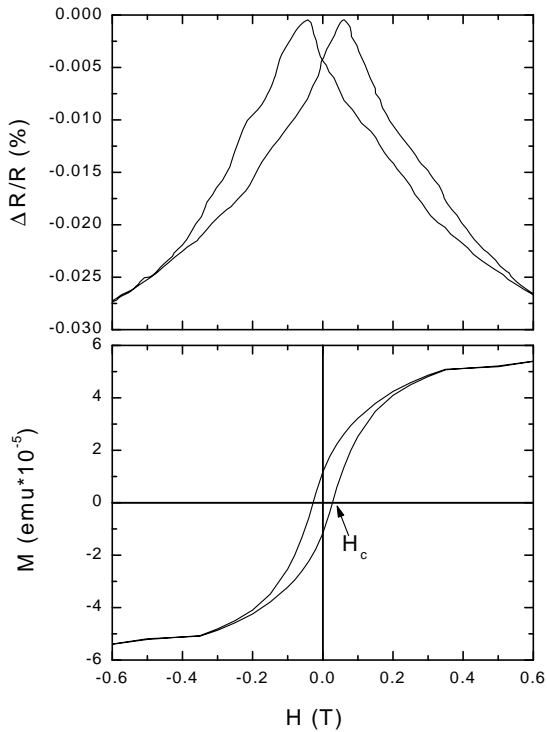}
}
\vspace{2mm}
\narrowtext
\caption{ MR and M-H curves taken at T=4.2K for a 23{\AA} 
 thick Ni film evaporated at room temperature . The arrow marks the position of the coercive field, $\text{H}_{c}$.}
\end{figure} 

We quantify the degree of hysteresis in the system by studying the coercive field, $\text{H}_{c}$,
for which the magnetization in the system, M, equals zero. In a granular film this is the 
field required to totally randomize the magnetic orientations, hence it should correlate
with the field at which the resistance peaks in the MR. Figure 4 compares the MR
measurements and measurement of the magnetization versus field (M-H) hysteresis loop performed 
in a SQUID magnetometer. Though  $\text{H}_{c}$ differs slightly 
in these two experiments (the differences will be discussed in length elsewhere \cite{future})
the behavior as a function of temperature or film thickness is similar.
The dependence of $\text{H}_{c}$ on sheet resistance 
for the different overlayered materials, depicted in figure 5, clearly demonstrates 
that there is a correlation between the coupling strength  of
the medium and its magnetic susceptibility. Pd, which is a strong paramagnet
($\chi\approx$11$\cdot$$10^{-6}$emu/gr$\cdot$Oe) has an effect which is 
nearly as strong as that of adding Ni itself.
Ti, a weaker paramagnet ($\chi\approx$1.5$\cdot$$10^{-6}$emu/gr$\cdot$Oe), has a significant but smaller coupling
coefficient. 
Diamagnetic materials, such as Ag or Cu \cite{rem2}, also couple between the Ni grains, though the 
influence of diamagnetic overlayers is much weaker than that of paramagnets. 
In contrast, Ge, which is an 
insulator, does not induce any magnetic coupling at all. Notice that we have
deposited a Ge overlayer which is much thicker than that of the metallic cases. 
Nevertheless, Ge seems to have no impact on the MR curve shape, in particular, it does 
not generate hysteretic behavior.
\begin{figure}[tb]
\centerline{
\epsfxsize=90mm
\epsfbox{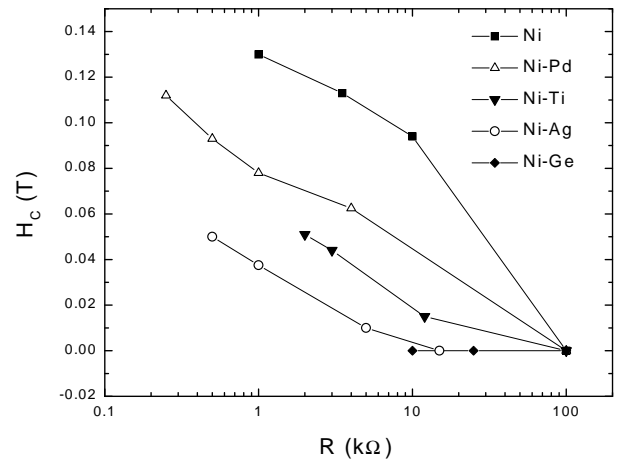}
}
\vspace{2mm}
\narrowtext
\caption{Coercive field at T=4.2K as a function of sheet resistance for
the different overlayer materials.}
\end{figure}
  We have considered a number of possible scenarios for magnetic coupling through the
normal medium. Classical magnetostatic coupling (dipole-dipole interaction 
between the grains) can be ruled out since the grains are isolated to begin with, 
hence, the magnetic interaction is obviously too weak to cause grain coupling. A strong
candidate for the coupling mechanism is an exchange interaction between the grains 
mediated by the conduction electrons in the metal. Such behavior is seen
in many magnetic heterostructures where magnetic layers are separated by non-magnetic
spacers. In these systems it is found that the magnetic layers couple ferromagnetically
or anti-ferromagnetically depending on the spacer thickness \cite{cu-exp}. This behavior has
been attributed to an RKKY-like
interaction \cite{RKKY,bruno}  which is oscillatory in space with a period of 1/2$\text{k}_{f}$. There are a number 
of problems in trying to ascribe this mechanism to the results in our case. 
In the first place the 
RKKY process is very sensitive to the spacing between the magnets. Calculations
show \cite{bruno} that thickness roughness on the order of atomic spacing reduces the amplitude of short 
period oscillations by more than an order of magnitude. In our granular film there is no single
grain spacing and the grains themselves are far from being atomically smooth, hence,
the effect of an RKKY interaction is expected to average out. Furthermore, RKKY coupling 
is a Coulomb effect which depends mainly on the Fermi surface details and not on the 
magnetic properties of the spacing layer. 
There is no apparent reason why Ti should induce stronger coupling than Ag. In fact,
Cu, which is the prototype spacer in multilayer systems, has hardly any effect
in our granular samples. The fact that we observe a 
strong correlation of hysteresis with magnetic susceptibility implies that the relevant mechanism  
is one in which a magnetic 
moment is induced in the intermediate medium which, in turn, couples the 
magnetic grains. We note the similarity to superconductivity, where the  proximity
effect is enhanced for 
a normal material which is characterized by an internal strong electron-phonon coupling, such as a superconductor
above its $\text{T}_{C}$. 

In conclusion we have shown that isolated magnetic grains can be coupled via an
intermediate medium which is not magnetic, resulting in a superparamagnetic-ferromagnetic 
crossover. This effect is stronger for paramagnetic 
media but exists also for diamagnets. We observe a clear relation between the coupling 
strength and the magnetic properties of the overlayer medium.  This phenomenon bears a striking
resemblance to an analogous experiment performed on granular superconductors  in
which an overlayer of Ag proximity couples isolated superconducting 
Pb grains\cite{Lynne}. The proximity coupling is expected to decay exponentially
with the distance between grains. Within the framework of the current experiment
we cannot yet provide an accurate evaluation of the relevant coherence length. 
Extraction of a coherence length will be pursued in future work. However,
a rough estimation of the distance between our 
grains ($\approx$10{\AA}) points towards relatively short coherence lengths, 
especially in the diamagnetic layers. 
Hence, in high quality
multilayers, the proximity effect can be expected to dominate for short scales 
(depending on the spacer magnetic properties)
and the magnetic layers will couple ferromagnetically.  For thicker spacers an RKKY- like 
exchange process, which decays like a power law, may set in and the 
sign of the coupling will oscillate with the thickness.   

We are grateful for technical help we received from T. Kirk and for illuminating discussions with D. Arovas, 
A.M. Berkowitz, F. Hellman, S. Sankar and H. Suhl. This research was supported by AFSOR
grant no. F49620-92-j-0070

\end{multicols}
\end{document}